%% file: bls_final.tex
\documentclass[11pt,letterpaper]{llncs}
\usepackage{amsmath}
\usepackage{amssymb}
\usepackage{graphicx}
\usepackage{marvosym}
\usepackage{url}
\usepackage{color, colortbl}
\usepackage{xcolor}
\usepackage{fancyhdr}
\usepackage{pdfpages}
\usepackage{multirow}

\usepackage{geometry}
\newcommand{\keywords}[1]{\par\addvspace\baselineskip
\noindent\keywordname\enspace\ignorespaces#1}

\fancypagestyle{firstpage}{\fancyhf{}
\rfoot{\thepage}
}

\begin{document}
\input{macro}


\title{\LARGE{A Study to Evaluate the Impact of \lora{} Fine-tuning on the Performance of Non-functional Requirements Classification}}


%
%
\author{\Large{Xia Li, Allen Kim}}
\institute{\large{The Department of Software Engineering and Game Design and Development,\\ Kennesaw State University,\\Marietta, USA}}

%


%
%

\maketitle
\thispagestyle{firstpage}

\begin{abstract}
Classifying Non-Functional Requirements (NFRs) in software development life cycle is critical.  Inspired by the theory of transfer learning, researchers apply powerful \pretrain{}s for NFR classification. However, full fine-tuning by updating all parameters of the \pretrain{}s is often impractical due to the huge number of parameters involved (e.g., 175 billion trainable parameters in GPT-3).
In this paper, we apply Low-Rank Adaptation (LoRA) fine-tuning approach into NFR classification based on prompt-based learning to investigate its impact. The experiments show that \lora{} can significantly reduce the execution cost (up to 68\% reduction) without too much loss of effectiveness in classification (only 2\%-3\% decrease). The results show that \lora{} can be practical in more complicated classification cases with larger dataset and \pretrain{}s. 

\end{abstract}

\section{Introduction}
Non-Functional Requirements (NFRs) in the whole software development life cycle are critical for meeting users' expectations and ensuring the system performs well across various dimensions, such as quality, usability, security, and performance. Unlike functional requirements that focus on the specific actions of the system, NFRs define the overall behavior and constraints of the system, which makes them essential in guiding architectural design decisions. For example, NFRs help determine how secure, scalable, or reliable the system needs to be. Given their importance, it is essential to identify and extract NFRs from software requirements specification (SRS) documents early in the development process to ensure that they are adequately addressed.  However, categorizing NFRs is a time-consuming task due to the complexity of their descriptions with natural languages. Consequently, the process is error-prone and requires a significant effort to ensure that all NFRs are properly understood and incorporated into the system design.

In the past decades, various machine learning and deep learning techniques have been applied to classify Non-Functional Requirements. For example, EzzatiKarami et al.~\cite{ezzatikarami2021automatically} used Support Vector Machine 
 (SVM) and Decision Tree for NFR classification by combining different feature extraction techniques. Rahimi et al.~\cite{rahimi2021one} used four deep learning models (e.g., LSTM, BiLSTM, GRU, and CNN) for accurate requirements classification. With the development and application of transformer models~\cite{vaswani2017attention}, various powerful \pretrain{}s (e.g., BERT~\cite{devlin2018bert}, GPT~\cite{brown2020language}) have been applied into requirements classification inspired by the theory of transfer learning. The basic idea is that the pre-trained foundation models can be trained in advance and users can fine-tune the models for their specific downstream tasks. For example, Hey et al.~\cite{hey2020norbert} proposed NoRBERT to fine-tune the BERT model and apply it to different tasks for requirements classification. Luo et al.~\cite{luo2022prcbert} proposed PRCBERT by applying prompt engineering for requirement classification using BERT model with flexible prompt templates. Despite the promising performances of current studies with \pretrain{}s, the major issues is the cost to fine-tune the models. Currently, there are huge amount of parameters (up to billions) to be tuned for modern large language models, indicating that a lot of time and resources will be spent to fine-tune better results for the specific task. To overcome the problem, a parameter-efficient fine-tuning (PEFT) approach called Low-Rank Adaptation (\lora{}~\cite{hu2021lora}) is proposed to accelerate the fine-tuning of large models while consuming less memory.  In this paper, we propose to conduct an extensive study to evaluate the impact of \lora{} on NFR classification. We apply p-tuning~\cite{liu2023gpt} which is a prompt-based approach on various \pretrain{}s such as BERT and \roberta{}~\cite{liu2019roberta}. The reason to use \ptuning{} is that it can provide stable performance for classification tasks by designing learnable templates. Our study indicates that \lora{} can significantly reduce the execution cost (e.g., up to 68\% reduction) without too much decrease of classification abilities. 

The structure of the paper is as follows. In Section \ref{sec:work}, we introduce the studies related to software requirements classification. In Section \ref{sec:approach}, we illustrate the approach of our study. In Section \ref{sec:design} and Section \ref{sec:results}, we discuss our experimental configurations as well as the results analysis, respectively. We discuss the threats to validity in Section \ref{sec:threats} and conclude the paper in Section \ref{sec:conclusion}.

\section{Related Work}
\label{sec:work}

\subsection{Traditional learning techniques for requirements classification}
In recent years, machine learning (ML) and deep learning (DL) techniques have been increasingly applied to enhance software requirement classification.  For example, Canedo et al.~\cite{dias2020software} compare the performance of different models such as Logist Regression (LR), Support Vector Machine (SVM), Multinomial Naive Bayes (MNB) and K-Nearest Neighbors (KNN) for both functional and non-functional requirements. Amasaki et al.~\cite{amasaki2018effects} use vectorization methods (e.g., document embedding methods) with various supervised machine learning models for NFR Classification.   AlDhafer et al.~\cite{aldhafer2022end} use Bidirectional Gated Recurrent Neural Networks (BiGRU) to classify requirements by considering word sequences and character sequences as tokens.  Dekhtyar et al.~\cite{dekhtyar2017re} use Word2Vec embeddings and CNN for NFR classification.

\subsection{Pre-trained models for requirements classification}
Inspired by the theory of transfer learning, researchers seek to apply powerful \pretrain{}s into the filed of software requirements classification. Kici et al.~\cite{kici2021bert} fine-tune an existing pre-trained  model called DistilBERT to achieve promising performance compared with other deep learning methods such as LSTM and BiLSTM. Hey et al.~\cite{rahimi2021one} propose a new approach called NoRBERT to fine-tune the popular BERT model for requirements classification. 
Luo et al.~\cite{luo2022prcbert} apply prompt learning for requirements classification by using BERT model 
flexible prompt templates to achieve accurate requirements classification. Kaur et al.~\cite{KaurBERTCNN} present a new Bidirectional Encoder-Decoder Transformer-Convolutional Neural Network (BERT-CNN) model for requirements classification, performing better than the state-of-the-art baseline approach.

\subsection{\lora{} for classification tasks}
LoRA (Low-Rank Adaptation) fine-tuning is a technique used to efficiently adapt pre-trained large language models (LLMs) to specific tasks with minimal computational cost. Researchers have applied \lora{} into various fields for classification tasks to achieve promising performance.  Yang et al.~\cite{yang2024news} introduce an advanced methodology for financial news
topic classification by leveraging the Chatglm3-6b model~\cite{du2021glm} through \lora{} and Noise Enhanced Fine-Tuning (NEFT). Shuai et al.~\cite{shui2024emotion} introduces a method utilizing the Llama3-8b model for emotion text classification which is accelerated by \lora{} and FlashAttention techniques. McCleary et al.~\cite{mccleary2024tnm} explore the application of \lora{} fine-tuning approach with small language models for triple negative breast cancer cases with various large language models such as GPT 3.5, GPT 4 and Mistral’s 7B.  Aggarwal et al.~\cite{aggarwal2024advancing} apply LoRA to the realm of image classification for plant disease detection.

In this paper, we conduct an extensive study on the impact of \lora{} by fine-tuning various large language models such as BERT and \roberta{}. We apply \ptuning{}, which is a popular prompt-based learning technique through \pretrain{}s by designing learnable templates fed into the models for training.

\section{Study Approach}
\label{sec:approach}

\begin{figure}[h!]
\centering
\includegraphics[scale=0.48]{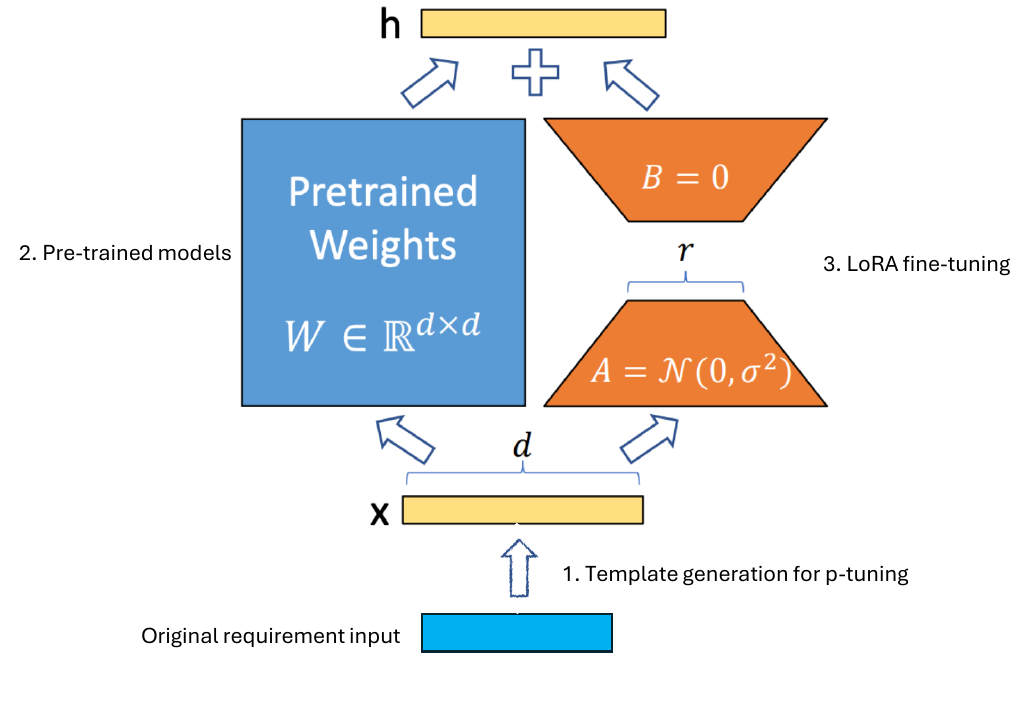}
\caption{Overview of the study approach}
\label{fig:overall}
\end{figure}

In this section, we introduce how we conduct the experiments for NFR classification based on \ptuning{} and \lora{} fine-tuning.
The overall process is shown in Figure~\ref{fig:overall}, including following steps. Please note that we cite the famous illustration of \lora{}~\cite{hu2021lora} as the fine-tuning part in the figure.
First, we convert the original requirement statements based on a specific template with learnable tokens based on \ptuning{}.
Second, the new template is fed into different \pretrain{}s to predict the category of requirement. Third, during the training process, we apply \lora{} approach to freeze the weights of original \pretrain{} by adding an alternative matrices. The target label can be predicted through back propagation with updated weights. Next, we introduce the basic ideas of \ptuning{} and \lora{} used in our study.

\subsection{P-tuning}
\label{sec:ptuning}

P-tuning is a prompt-based learning technique where certain learnable continuous prompt embeddings (soft templates) are created in concatenation with original input requirement sequence. The target label of the classification task is embedded in the sequence that \pretrain{}s can predict. For other techniques using \pretrain{}s, an additional neural network (e.g., RNN, CNN) is connected with the pre-trained models and \pretrain{}s are typically used to generate a final vector representation for input sequence with low correlation between the input sequence and the target task. However, \ptuning{} can let \pretrain{}s directly predict the masked token label to achieve better classification performance. For \ptuning{}, we use the following template as the input of the \pretrain{}s: \emph{[CLS][P][P][P]REQUIEMRNT SENTENCE[SEP]. [CLS][P][P][P]This requirement is related to [M][SEP]}. In the template design, [P] represents the learnable token without any real meanings while [M] is the masked token to represent the requirement category (e.g., performance, security, usability) that can be predicted by \pretrain{}s. [CLS] represents a special token  in the front of the original input text and [SEP] is a separator token to represent the segment of each sentence. Thus, one example of the final input can be `` \emph{[CLS][P][P][P] How well are the system and its data protected against attacks?[SEP]. [CLS][P][P][P]This requirement is related to [M][SEP]} ''.

\subsection{\lora{}}
\label{sec:lora}
To adapt current pre-trained large language models (e.g., GPT-3, LLama, and Mistral ) to specific tasks, fine-tuning is an essential step. However, full fine-tuning by updating all parameters of the \pretrain{}s is often impractical due to the huge number of parameters involved (e.g., 175 billion trainable parameters in GPT-3), making it computationally expensive and storage-intensive. To resolve this problem, \lora{} is proposed to offer a more efficient alternative by significantly reducing the number of trainable parameters. As shown in Figure~\ref{fig:overall}, \lora{} can freeze the weights of \pretrain{}s while insert two smaller matrices to represent the updated weights through low-rank decomposition. These new matrices can be trained to adapt to the data while keeping the overall number of changes low.  In detail, W0 with size $d \times d$ is the original weight matrix and kept unchanged during the training procedure. $\Delta$W is the new updated weights with the same size $d \times d$.  However, in \lora{}, a new parameter r is introduced to reduce the size of the matrix $\Delta$W.  The smaller matrices, A and B, are defined with a reduced size of $r \times d$, and $d \times r$. For \lora{}, only the weights of matrices A and B are trained to update during training process.  After the training process is completed, the new input x with a size of $1 \times d$ will be multiplied by both W and $\Delta$W, resulting in two d-sized output vectors. These vectors are then concatenated  element-wise to produce the final result, and the weights can be merged with the base model. The final output is a vector with size $h$ shown in the figure.  In this paper, we investigate how \lora{} can improve the efficiency of NFR classification and if the similar effectiveness can be kept with \lora{}.

\section{Experimental Design}
\label{sec:design}
In this section, we introduce the \pretrain{}s and dataset used in our study in Section 4.1 as well as the experimental configuration and evaluation metrics in Section 4.2.

\subsection{Pre-trained models and dataset}
\label{sec:models}

To evaluate \lora{} on \pretrain{}s with different sizes, we use four foundation models "bert-base-uncased" (110M parameters), "bert-large-uncased" (340M parameters), "roberta-base"(125M parameters) and "roberta-large"(355M parameters) that can be downloaded from the popular AI community Hugging Face\footnote{Hugging Face. https://huggingface.co/}.  
We use the widely used dataset PROMISE~\cite{sayyadpromise} with 914 non-functional requirements that consists of the following five pre-labeled categories: operability, maintainability, performance, security, and usability. To make the dataset adapted to the template,  we apply popular natural language processing steps (e.g., stemming, lemmatization, stop-word removal and conversion to lower case) via the widely used NLTK\footnote{NLTK toolkit. https://www.nltk.org/} toolkit to pre-process the dataset. We also remove special characters that are unique in different domains. 

\subsection{Experimental configuration and evaluation metrics}
\label{sec:config}

In our study, we split the original dataset into training set (80\%) and test set (20\%). We apply 10-fold cross-validation for the four \pretrain{}s. We use cross-entropy loss function since NFR classification is the classic multi-class classification problem.  We also use the popular evaluation metrics precision, recall and F1 score for classification problems. Their equations are as follows.
\[
\text{Recall} = \frac{TP}{TP + FN}
\]
\[
\text{Precision} = \frac{TP}{TP + FP}
\]
\[
F1 = 2 \times \frac{\text{Precision} \times \text{Recall}}{\text{Precision} + \text{Recall}}
\]
Where TP represents the number of True Positives, FP indicates the number of False Positives and FN is the number of False Negatives.

For the hyperparameters, we set the maximum input sequence length as 128, batch size as 8, learning rate as $3e^{-5}$, epochs as 32. To implement \lora{}, we use the pre-defined framework \textbf{peft} from Hugging Face. From the experiments, we can reduce the trained parameters to 20\% - 25\% from original \pretrain{}s. Also, we set the specific parameter rank r used in \lora{} as 8.  We also use AdamW optimizer \cite{loshchilov2017decoupled} in the training process.  All training and inference are executed on a server with Intel Core 13900K CPU, 32GB memory and NVIDIA RTX 4090 GPU.

\section{Results Analysis}
\label{sec:results}
\input{Results/Result1}
\input{Results/Result2}
In this section, we illustrate the impact of \lora{} on the efficiency and effectiveness of NFR classification with \ptuning{}, shown in Table \ref{tab:RQ1} and Table \ref{tab:RQ2}. In Table \ref{tab:RQ1}, the number indicates the execution cost of classification for 1-fold cross validation. From the results, we can find that \lora{} can significantly reduce the execution cost of NFR classification. For example, for the Besrt-large model, it can reduce the cost by 68.4\% (19 mins vs 6 mins). In Table \ref{tab:RQ2}, the first column represents the four models used in the study. The last three columns represent the values in terms of the three metrics: precision, recall and F1 score. Please note that the number outside the parentheses represents the value without \lora{} while the number in the parentheses indicates corresponding results with \lora{} fine-tuning. Also, all results are calculated as the average values of 10-fold cross-validation. From the results, we have following findings.   First, we can find that the two models from Roberta performs better than Bert models. For example, the F1 score of Bert-base is 81.92\% while it is 82.86\% for \roberta{}. The possible reasons could be that \roberta{} is trained on a larger corpus of text with robust representation of the language and it uses a dynamic masking strategy where different tokens are masked in each training example. Also, we can find that the large versions of both models perform better than their corresponding base model. It indicates that more weights trained for NFR classification can get better performance.  Third, with \lora{} fine-tuning approach, we can find that the performance are worse than corresponding models without \lora{}. However, the difference is only 2\%-3\%, showing that \lora{} can be practical and acceptable in more complicated classification cases with larger dataset and \pretrain{}s.

\section{Threats to Validity}
\label{sec:threats}

The main external threat to the validity is the dataset we used. In our study, we use the widely used data PROMISE for NFR classification to make our results as accurate as possible.
The internal threat to the validity is the implementation of the training and inference. To reduce the threat, we use the pre-defined framework of \lora{} from the popular AI community Hugging Face.

\section{Conclusion}
\label{sec:conclusion}

In this paper, we conducted a comprehensive study to evaluate impact of \lora{} fine-tuning approach on the performance of prompt-based non-functional requirements classification. Our experimental results show that with \lora{} fine-tuning approach, the execution cost can be reduced by up to 68\% and the performance are worse than corresponding models without \lora{}. However, the difference is only 2\%-3\%, showing that \lora{} can be practical and acceptable in more complicated classification cases with larger dataset and \pretrain{}s compared with the significant execution cost reduction.  In future, we plan to evaluate NFRs classification by using more fine-tuning approaches.

\vspace{2cm}



\end{document}

%% file: macro.tex
\definecolor{darkpastelgreen}{rgb}{0.01, 0.60, 0.24}
\definecolor{amber}{rgb}{0.7, 0.5, 0}

\newcommand{\Xia}[1]{{\color{blue}#1}}

\newcommand{\RE}{requirements engineering}
\newcommand{\stateart}{state-of-the-art}
\newcommand{\promptbase}{prompt-based}
\newcommand{\promptengin}{prompt engineering}
\newcommand{\pretrain}{pre-trained model}
\newcommand{\ptuning}{p-tuning}

\newcommand{\jdkshortseven}{JDK 1.7}
\newcommand{\jdklongseven}{JDK 1.7.0.80}

\newcommand{\jdkshorteight}{JDK 1.8}
\newcommand{\jdklongeight}{JDK 1.8.0.242}

\newcommand{\pit}{PIT}

\newcommand{\defectsj}{Defects4J}

\newcommand{\codeIn}[1]{\texttt{#1}}
\newcommand{\parabf}[1]{\noindent \textbf{#1}}

\newcounter{finding}
\newcommand{\finding}[1]{\refstepcounter{finding}
	\begin{mdframed}[linecolor=gray,roundcorner=12pt,backgroundcolor=gray!15,linewidth=1pt,leftmargin=0cm,rightmargin=0cm,topline=true,bottomline=true,skipabove=12pt]
		\textbf{Finding \arabic{finding}:} #1
	\end{mdframed}
}

\newcommand{\camreview}[1]{{\color{amber}[#1]}}
\newcommand{\camready}[1]{{\color{amber}[#1]}}
\newcommand{\old}[1]{{\color{red}[#1]}}

\newcommand{\susp}{Susp}
\newcommand{\tf}{n_f}   
\newcommand{\tp}{n_p}   

\newcommand{\ef}{n_f(e)}
\newcommand{\ep}{n_p(e)}
\newcommand{\nf}{n_f(\bar{e})}
\newcommand{\np}{n_p(\bar{e})}

\newcommand{\esf}{n_f(s)}
\newcommand{\esp}{n_p(s)}
\newcommand{\nsf}{n_f(\bar{s})}
\newcommand{\nsp}{n_p(\bar{s})}

\newcommand{\emf}{n_f(me)}
\newcommand{\emp}{n_p(me)}
\newcommand{\nmf}{n_f(\bar{me})}
\newcommand{\nmp}{n_p(\bar{me})}

\newcommand{\tpme}{n_p^{(m)}(e)}
\newcommand{\tfme}{n_f^{(m)}(e)}
\newcommand{\tfmn}{n_f^{(m)}(\bar{e})}
\newcommand{\tpmn}{n_p^{(m)}(\bar{e})}

\newcommand{\tpms}{n_p^{(m)}(s)}
\newcommand{\tfms}{n_f^{(m)}(s)}
\newcommand{\tfmns}{n_f^{(m)}(\bar{s})}
\newcommand{\tpmns}{n_p^{(m)}(\bar{s})}

\newcommand{\lora}{LoRA}
\newcommand{\roberta}{RoBERTa}

%% file: Results/Result1.tex
\begin{table}\center
\caption {Efficiency of NFR classification with and without \lora{}} 
\label{tab:RQ1}
\scalebox{1.2}{
\begin{tabular}{c|c|c|c}
\hline
Model&without \lora{}&with \lora{} & \%reduction \\

\hline

Bert-base&11 mins&5 mins&54.5\%\\
Bert-large&19 mins&6 mins&68.4\%\\
Roberta-base&13 mins&7 mins&46.2\%\\
Roberta-large&20 mins&9 mins &55\%\\

\hline

\end{tabular}}
\end{table}

%% file: Results/Result2.tex
\begin{table}\center
\caption {Effectiveness of NFR classification with and without \lora{}} 
\label{tab:RQ2}
\scalebox{1.2}{
\begin{tabular}{c|c|c|c}
\hline
Model&Precision&Recall&F1 score\\

\hline

Bert-base&81.39\%(78.75\%)&82.46\%(78.13\%)&81.92\%(78.40\%)\\\
Bert-large&82.37\%(79.26\%)&82.90\%(79.38\%)&82.44\%(79.32\%)\\
Roberta-base&82.57\%(80.73\%)&83.17\%(79.75\%)&82.86\%(80.26\%)\\
Roberta-large&83.45\%(81.29\%)&83.57\%(80.15\%)&83.42\%(80.78\%)\\

\hline

\end{tabular}}
\end{table}